\documentclass[aps,prl,twocolumn]{revtex4}

\usepackage{textcomp}
\usepackage{graphicx}
\usepackage{latexsym}
\usepackage{amsmath}
\usepackage{amsfonts}
\usepackage[utf8]{inputenc}

\begin{document}
\title{Experimental search for one-dimensional edge states at surface steps of the topological insulator Bi$_2$Se$_3$:  Distinguishing between effects and artifacts}
\author{N.I. Fedotov}
\author{S.V. Zaitsev-Zotov}
\affiliation{Kotel'nikov Institute of Radioengineering and Electronics of RAS, Mokhovaya 11, bld. 7, Moscow, 125009 Russia}

\date{\today}
\begin{abstract}
The results of a detailed study of the topological insulator Bi$_2$Se$_3$ surface state energy structure in the vicinity of surface steps using scanning tunneling microscopy  and spectroscopy methods are presented. An increase in the chemical potential level $\mu$ near the step edge is observed. The value of the increase $\delta \mu\sim 0.1$~eV is found to correlate with the step height. The effect is caused by redistribution of electron wave functions between the outer and inner edges of surface steps, as  known for normal metals. The smaller value of the chemical potential shift and its larger characteristic length of $\sim 10$~nm reflect specifics of the helical surface states. This increase is accompanied by enlargement of the normalized differential tunneling conductance in the helical surface states energy region and thereby produces the illusion of the appearance of edge states. We show that the enlargement is reproduced in the framework of the tunneling model taking into account the tunneling gap transparency change when the chemical potential moves away from the Dirac point.
\end{abstract}
\maketitle

\section{Introduction}

Termination of the periodic potential of a crystal by a surface results in the appearance of surface states known as Tamm \cite{Tamm} or Shokley \cite{Shockley} states. 
The states are degenerate by spin, localized near the surface and decay exponentially in the direction perpendicular to the surface. The recent classification of crystalline solids by topological invariant revealed another type of surface state whose existence is protected by time-reversal symmetry \cite{TI}. These gapless helical Dirac fermion states have a specific conelike spectrum (the so-called Dirac cone) crossing the entire bulk energy gap, they are nondegenerate (except for the cone apex, called the Dirac point) and are characterized by the spin-momentum locking, with the spin being orthogonal to the momentum. In particular, the three-dimensional topological insulators (TIs) Bi$_2$Te$_3$ and Bi$_2$Se$_3$ host a single Dirac cone with the Dirac point at the $\Gamma$ point of the Brillouin zone \cite{BiTeBiSe}.

The crystal surface is also a periodic object, and a question arises about what happens to the surface states if this periodicity is also terminated, for instance, by a crystal face at a surface step? In the case of topologically trivial material with conducting surface states the answer is well known: an interference pattern appears due to interference of incident and scattered states \cite{interfer1,interfer2,Simon}. The answer for topologically nontrivial insulators is still not certain. If an insulator is characterized by nonzero topological invariant $Z_2$, then the existence of the surface states is protected by the time-reversal symmetry and the momentum of the surface electrons is locked to their spin. A change in momentum direction now means a spin direction change. This circumstance destroys the interference and new features are expected to appear. 

Scanning tunneling microscopy (STM) and spectroscopy (STS) provide the most direct information on local properties in the vicinity of various surface defects. Experimental studies of the effect of surface steps on surface electronic states in TIs  using these methods revealed a number of new features. First of all,  standard interference patterns formed around the steps  are observed for electronic states far from the Dirac point ($k\gtrsim 0.1$~\AA$^{-1}$) in  both  Bi$_2$Te$_3$ \cite{Zhang,Alpichshev} and Bi$_2$Se$_3$ \cite{Song}.
The interference pattern amplitude vanishes upon approaching the bulk energy gap edge from the bulk band state side, so no interference is seen for pure helical states \cite{Alpichshev}. 

Electron states which may appear at the edge between two adjacent TI crystal surfaces hosting helical states (or surfaces with different velocities of masless Dirac fermions) were also studied theoretically \cite{Deb,Paananen,Takane,Biswas,Lee,Seshardi,Sen}. Edge states were predicted to appear along the edge between two surfaces  \cite{Deb,Sen} or at the sides of a strip \cite{Paananen} of a three-dimensional topological insulator such as Bi$_2$Se$_3$ but, to the best of our knowledge, heve not yet been reported experimentally for this material. 
In the topological insulator Bi$_2$Te$_3$ a bound state along a step on the surface was found to appear \cite{Alpichshev}. Namely, the local density of states increases by a few tens of percent within 1-2 nm near the step edge, forming thereby a sort of an edge state \cite{Alpichshev}. Energy dispersion of such states and their existence in other materials such as Bi$_2$Se$_3$ are still  open questions.

Here we present the results of a detailed study of the Bi$_2$Se$_3$ energy structure in the vicinity of surface steps using STS methods. We observe a smooth variation of the chemical potential level by 0.1--0.2~eV over a distance of $\sim 10$~nm. In addition, an increase in the normalized differential tunneling conduction in the vicinity of the steps is observed and produces an illusion of the edge states. We show here that this increase can be practically entirely accounted for if the bias-induced change in the transparency of the tunneling gap is taken into account. 
\section{Experimental and methodical notes}

Bi$_2$Se$_3$ crystals were grown from a mixture of Bi and Se with a 3\% excess of Se over  stoichiometric quantity
 by heating followed by  smooth cooling down in evacuated quartz ampules; growth details are available elsewhere \cite[]{Dmitriev}. The experiments were carried out with ``standard'' {\em n}-type  Bi$_2$Se$_3$ crystals (group I), as well as on  Bi$_2$Se$_3$ crystals with the chemical potential position inside the bulk band gap (group II).  STM and STS measurements were performed with an  Omicron LT-STM operating at a base pressure of $2\times10^{-11}$ Torr.  The samples were exfoliated {\em in situ} at room temperature and transferred to the low-temperature section of the STM kept at liquid helium temperature. Tips cut from  Pt-Rh wire were used for imaging and spectroscopy. The quality of tips was checked on Au foil before and after the measurements by checking for linearity of the $I$-$V$ characteristic in the vicinity of  $V=0$. If needed, we performed a tip recovery procedure which included briefly dipping the tip into the Au foil followed by the standard tip control procedure described above. The STM images were recorded in the constant current mode. $I$-$V$ curves were acquired in the spectroscopy mode and numerically differentiated. The results discussed below represent a typical behavior observed in 2 group I samples (six steps) and seven group II samples (14 steps). Some preliminary results obtained for group II samples can be found in Ref.~\cite{Dmitriev}
 
Structurally, Bi$_2$Se$_3$ consists of quintuple layers (QL) Se-Bi-Se-Bi-Se, stacked on top of each other and bound together by van der Waals forces, so the crystals are easily cleaved between the layers and steps on the Bi$_2$Se$_3$ (111) surface usually correspond to an integer number of QLs.

\begin{figure}
\includegraphics[width=0.44\textwidth]{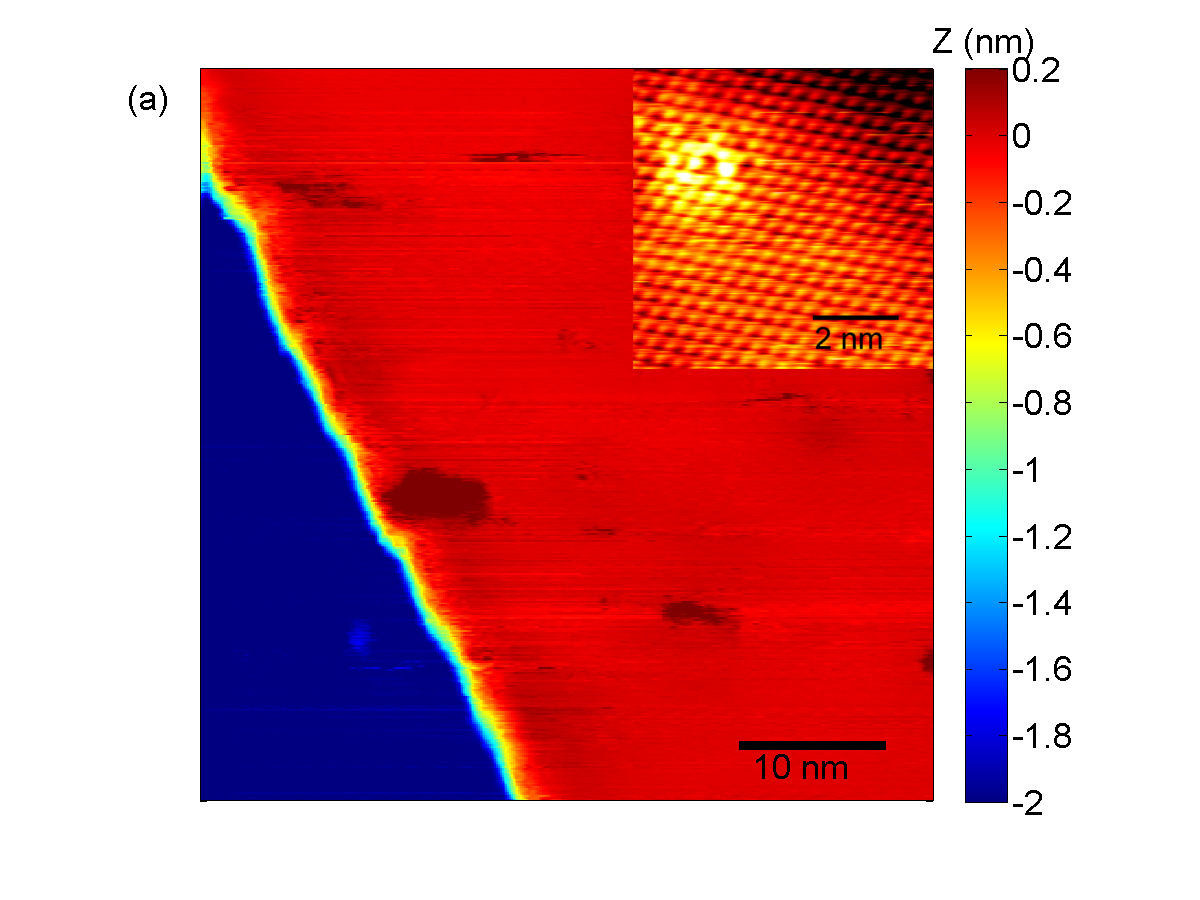}
\includegraphics[width=0.44\textwidth]{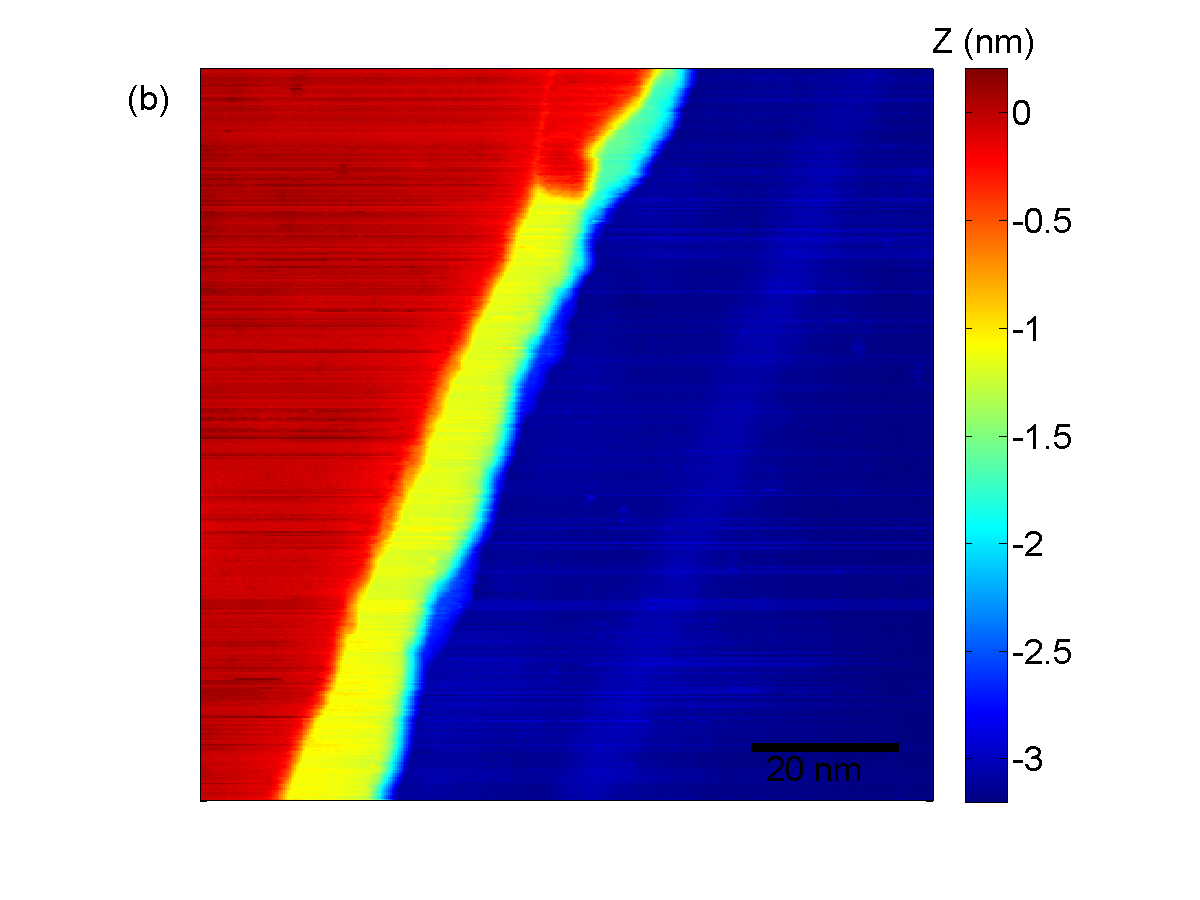}
\caption{STM images of Bi$_2$Se$_3$ surfaces with steps of different heights: (a) Group I sample with 2QL step; the inset shows the atomic resolution image. (b) Group II sample with 1QL and 2QL steps. In both cases $V_t=-0.4$~V, $I_t=100$~pA, $T=5$~K. }
\label{surface}
\end{figure}

Regions of the sample surface containing steps and other extended defects were selected for the study. Typical STM images of such regions  are shown in Fig.~\ref{surface}. Terraces separated by one [Fig.~\ref{surface}(a)] or two steps [Fig.~\ref{surface}(b)] are clearly seen. Step height values correspond to one or two quintuple layers. The inset in Fig.~\ref{surface}(a) shows a fragment of the surface with atomic resolution. We see that the step edge consists  mostly of $[\overline 2110]$ segments connected by relatively short $[\overline 1100]$ ones. 

\begin{figure}
\includegraphics[width=0.4\textwidth]{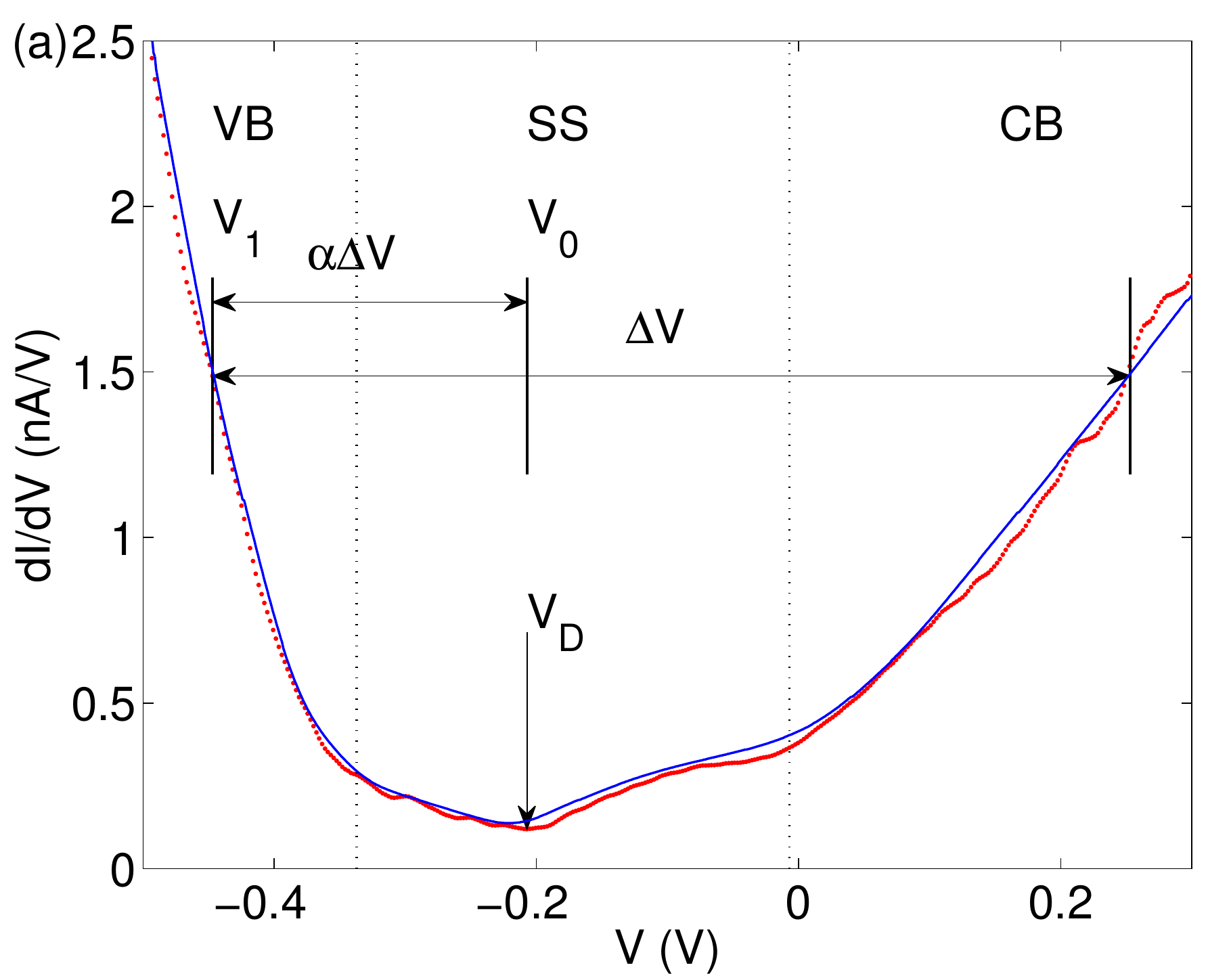}
\includegraphics[width=0.42\textwidth]{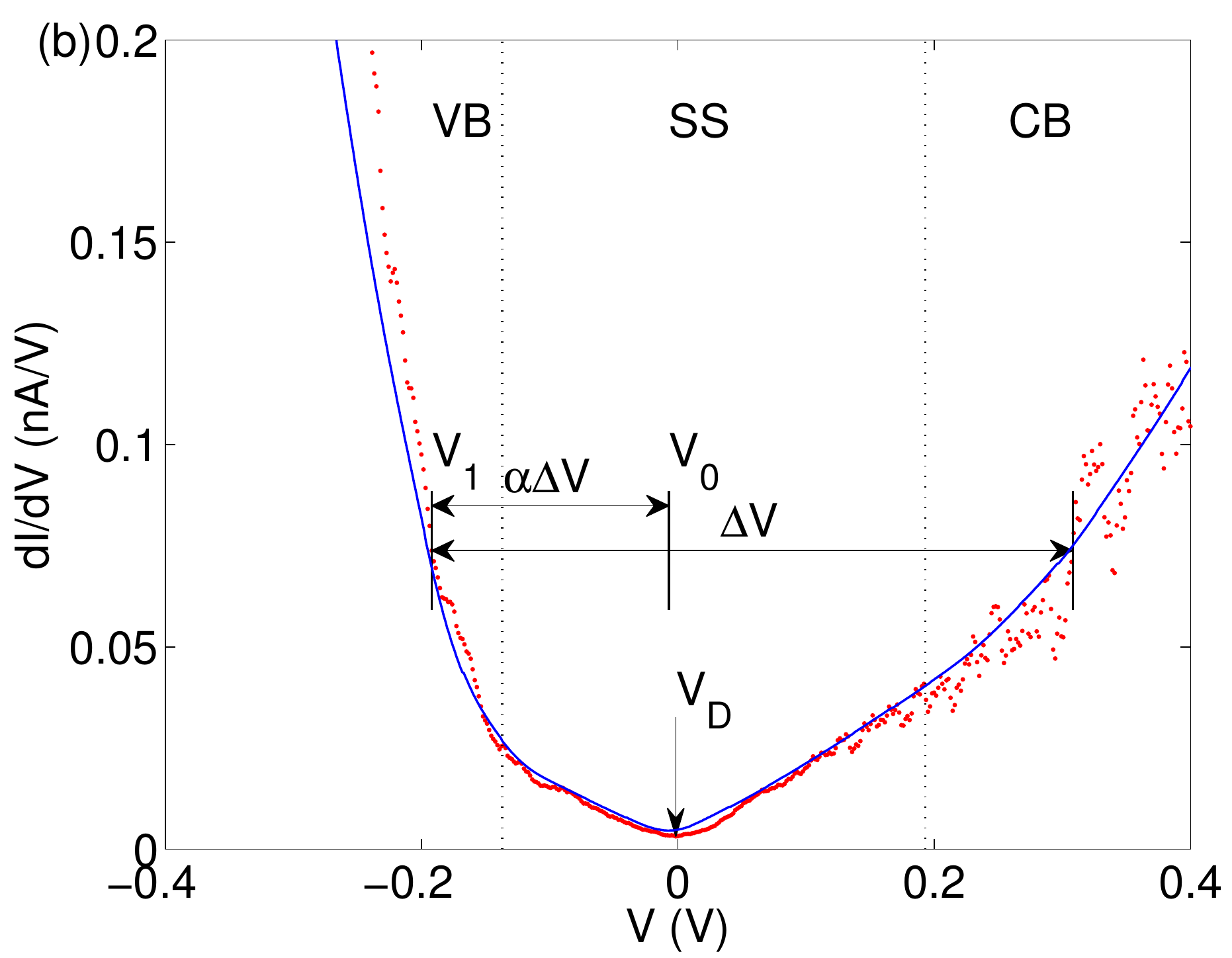}
\caption{Typical differential  tunneling conductance curves on the Bi$_2$Se$_3$ (111) surface  away from defects. The bulk valence band (VB), bulk conduction band (CB) and  surface states (SS) are separated by vertical lines. The arrow points to the Dirac point identified as the minimum of the differential conductance curve.  $V_1$ corresponds to (a) $\Delta V=0.7$~V for group I samples and (b) 0.5~V for group II samples (see text for details). Set point $V_t=-0.4$~V, $I_t=100$~pA, $T= 5$~K. Solid lines show $dI/dV$ curves calculated by using Eqs.~(\ref{IV}) and (\ref{T}) \cite{comment} with the model DOS shown in the inset in Fig.~\ref{didv}.}
\label{vah}
\end{figure}

A typical differential tunneling conductance curve obtained away from defects is shown in Fig.~\ref{vah}(a). 
The Dirac point of Bi$_2$Se$_3$ is in the bulk band gap and is identified as the minimum of the V-shaped feature. Identification of the bulk valence and conduction band positions is based on the assumption of a bulk energy gap value of 0.33~eV. 
The chemical potential level of group I samples is near the bottom of the bulk conduction band providing {\em n}-type conduction, [Fig.~\ref{vah}(a)], whereas it is within $\pm 0.05$ eV of the Dirac point in group II samples [Fig.~\ref{vah}(b)]. 

We want to analyze the spatial variation of the local density of states (LDOS) obtained from scanning tunneling spectra.
So a proper choice of data normalization is required. 
If the chemical potential varies along the sample surface then normalization of $dI/dV$ to a fixed set point is definitely not a good choice since this set point will correspond to different values of the LDOS. Here we normalize the differential conductance curves by their values at a selected position in the energy structure of the sample. Our choice for this position is the following. We assume that the bulk energy structure (namely,  separation of valence and conduction bands) does not change near the step. We select a certain $\Delta V$  
exceeding the bulk energy gap and find  $V_1$  at which the energy width of the differential conductance curve equals to $\Delta V $ (see Fig.~\ref{vah}), so that  $\frac{dI}{dV}|_{V_1} = \frac{dI}{dV}|_{V_1+\Delta V}=G_t$. Then $G_t$ is used for normalization of $dI/dV$ curves. 

$V_1$ allows us to trace the shift of the bulk bands, and therefore is a measure of the electrostatic potential. In its turn, the position of the Dirac point of the surface states can be obtained as the voltage $V_D$ corresponding to  $\min(dI/dV)$. To compare one to the other  it is more convenient  to use  $V_0=V_1+\alpha\Delta V$  instead of $V_1$, where $\alpha$ is chosen to provide $V_0=V_D$ far from defects.
For group I samples we choose  $\Delta V = 0.7$~V, $\alpha = 0.34 $ and for group II samples we choose $\Delta V = 0.5$~V, $\alpha = 0.37$ since the voltage range in group II measurements is not wide enough for $\Delta V = 0.7$ V.

\section{Results}
\subsubsection{STS results along a line. Shift of the chemical potential level}

\begin{figure}
 \ \ \ \ \ \ \includegraphics[width=0.43\textwidth]{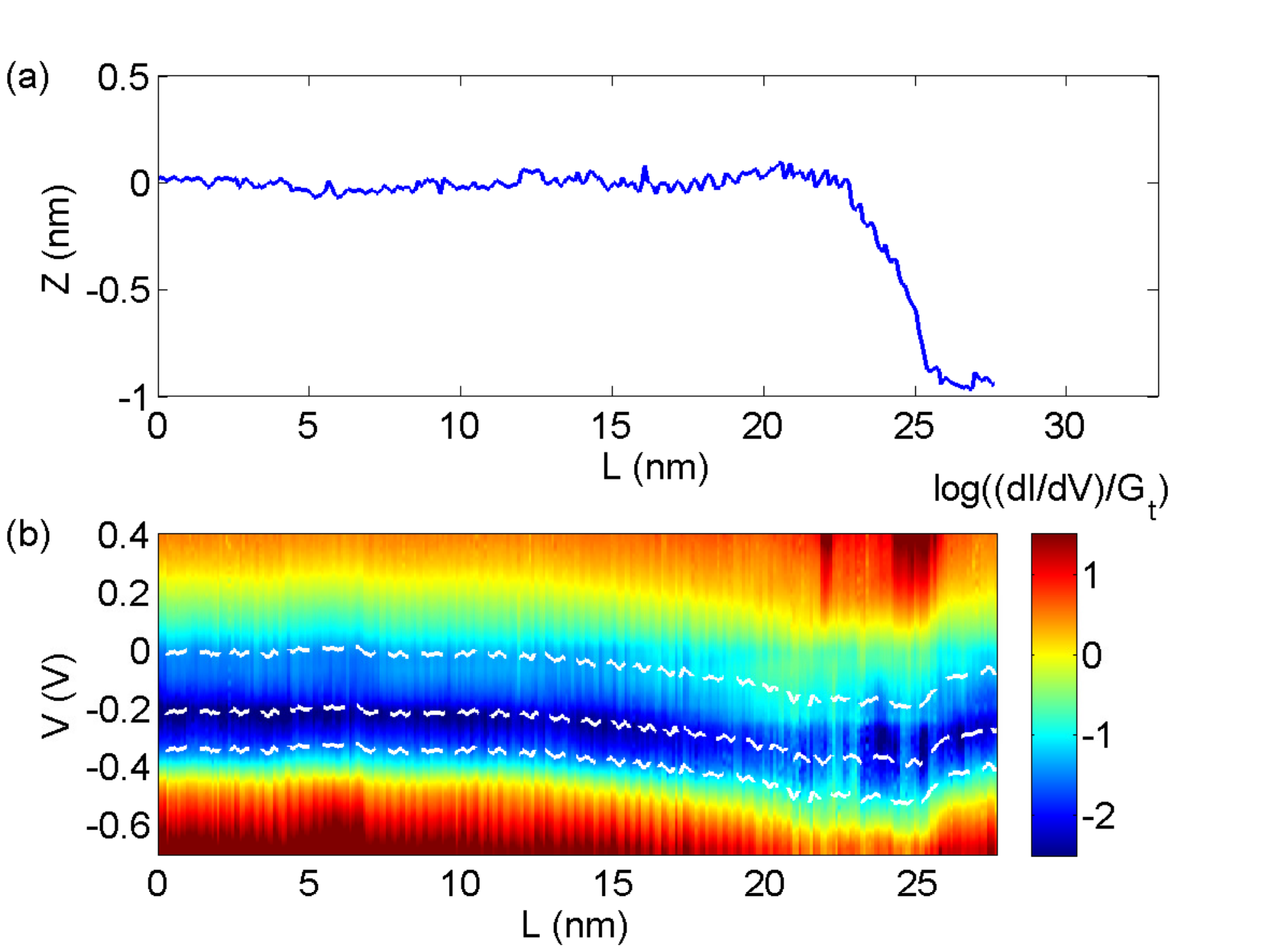}
\includegraphics[width=0.4\textwidth]{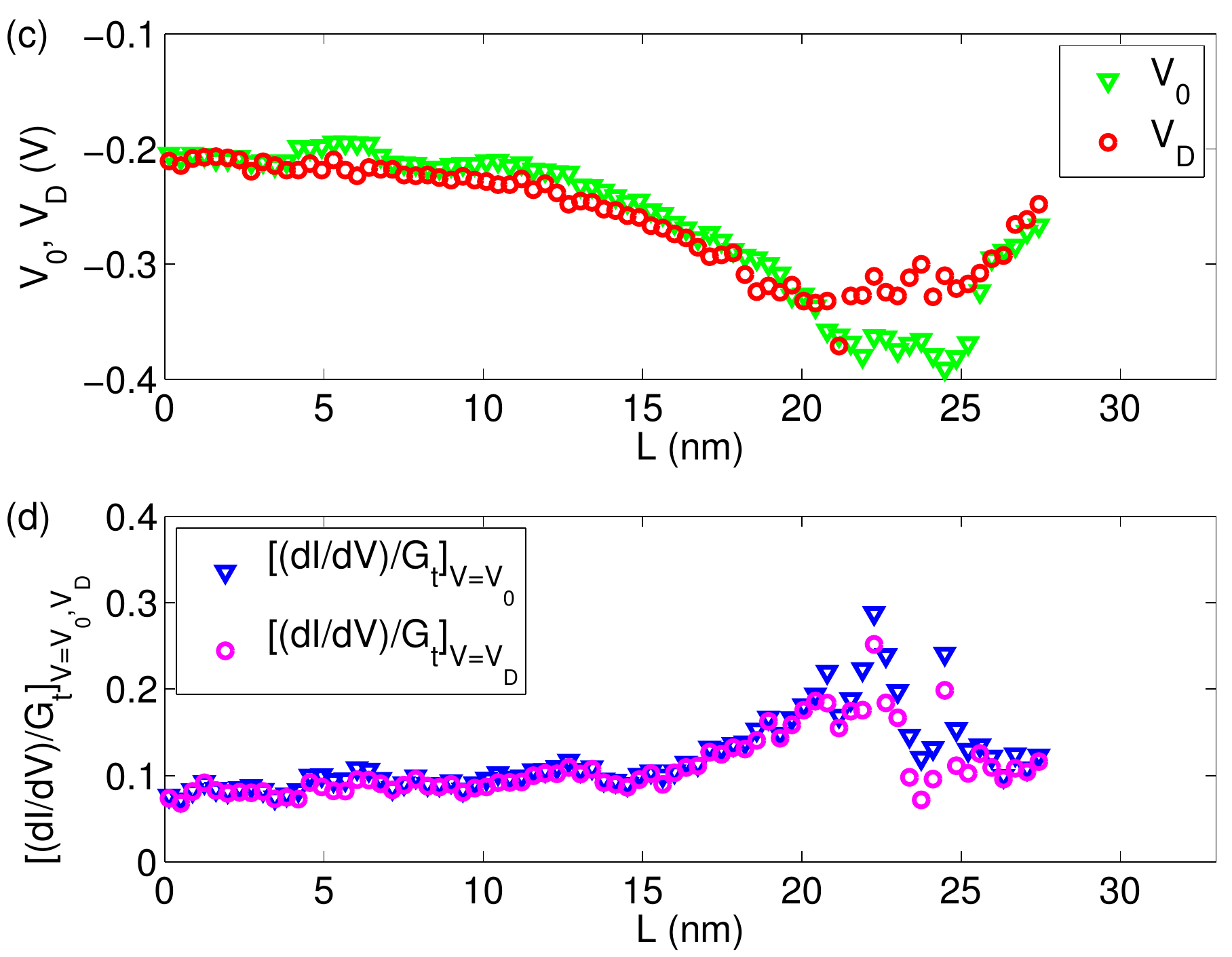}
\caption{(a) Profile along a line crossing the 1QL step on the Bi$_2$Se$_3$ (group I) surface. (b) The respective sets of $dI/dV$ curves, positions of VB and CB edges, and $V_0$  shown by dashed white lines, pay attention to the logarithmic color scale). (c) $V_D$ and $V_0$, and (d) $[(dI/dV)/G_t]_{V=V_0}$ and $[(dI/dV)/G_t]_{V=V_D}$ along the  line. $ I$-$V$ curves were collected at $V_t = -0.4$~V, $I_t = 100$~pA, $T=5$~K.}
\label{line1}
\end{figure}

\begin{figure}
\ \ \ \ \includegraphics[width=0.45\textwidth]{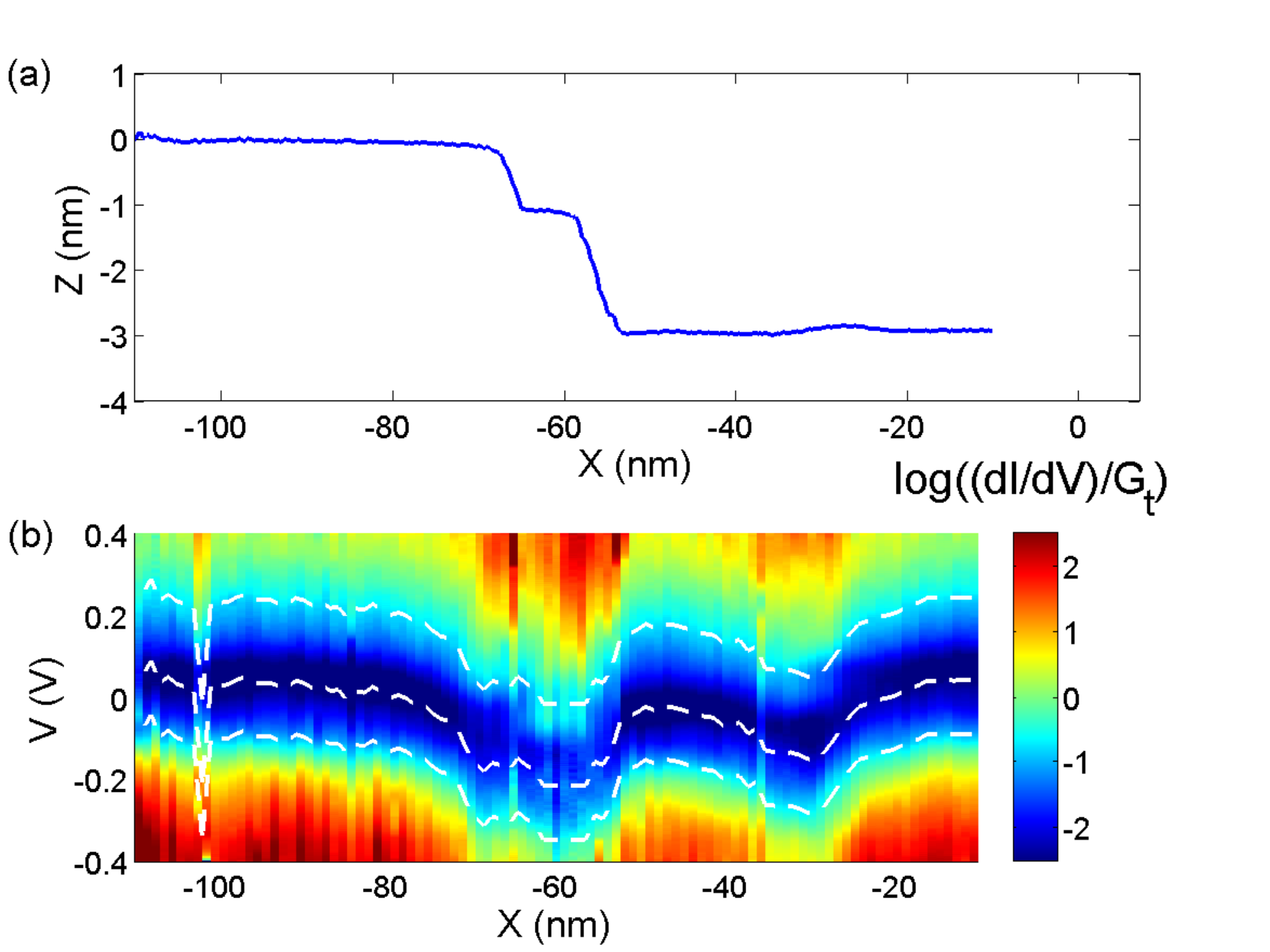}
\includegraphics[width=0.4\textwidth]{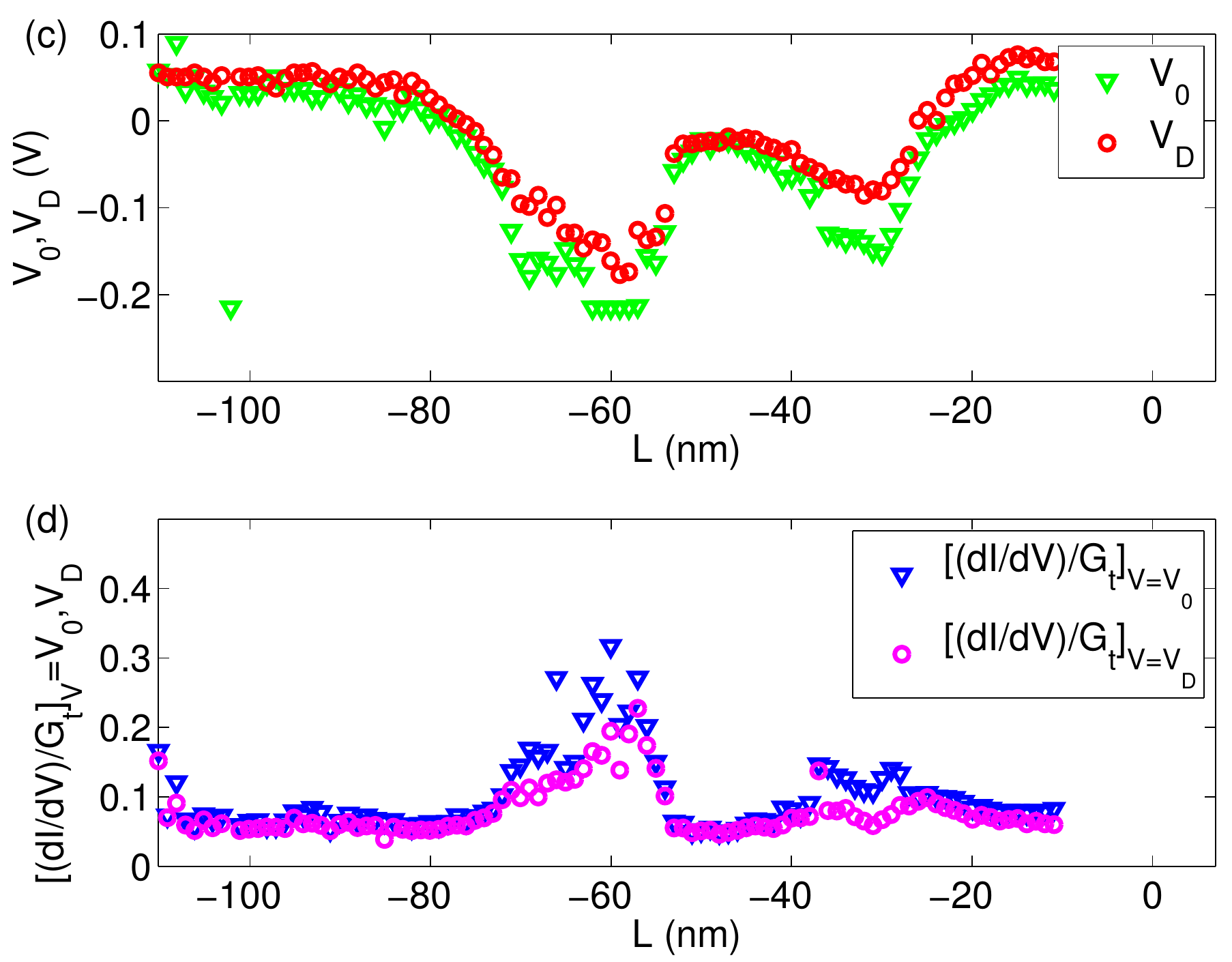} \ \ \ \ \
\caption{Profile along a scan line crossing 1QL  and 2QL steps on Bi$_2$Se$_3$ (group II) surface (a); respective sets of $dI/dV$ curves, positions of VB and CB edges and $V_0$ are shown by dashed white lines ((b), pay attention to the logarithmic color scale). (c) $V_D$ and $V_0$, and (d) $[(dI/dV)/G_t]_{V=V_0}$ and $[(dI/dV)/G_t]_{V=V_D}$ along the scan  line. $ I$-$V$ curves were collected at $V_t = -0.4$~V, $I_t = 100$~pA, $T=5$~K.}
\label{line2}
\end{figure}

Figure~\ref{line1}(a) shows the surface profile along a line crossing a step on a group I sample.
The step edge is at $L\approx 22$~nm, and the step height corresponds to 1QL. 
The step-edge profile looks smooth due to the finite radius of the tip ($\approx 5$~nm in this particular case).

Using a set of {\em I-V} curves taken in 300 equally separated points along the line (each {\em I-V} curve is an average of 20 independent curves taken at the same point) we find the spacial distribution of the normalized $dI/dV$ curves 
[Fig.~\ref{line1}(b)].   A shift of the chemical potential near the step edge is apparent. Namely, the $dI/dV$ curves move as a whole towards the bulk valence band by  $\approx 0.15$~eV while approaching  the step and restore their initial position on the lower terrace. 

From the same set of {\em I-V} curves we calculate the $V_0(L)$, $V_D(L)$, $[(dI/dV)/G_t]_{V=V_0}$ and $[(dI/dV)/G_t]_{V=V_D}$ dependences  [Fig.~\ref{line1}(c,d)]. We see that both $V_0(L)$  and $V_D(L)$ start to shift at a distance $\delta l\approx 10$~nm from the step edge (Fig.~\ref{line1}(c)). The shift corresponds to a positive charge accumulated near the step edge and is accompanied by an increase in the normalized differential tunneling conductance at the $V_0$ and  $V_D$ positions [Fig.~\ref{line1}(d)]. 

Very similar features are observed in group II samples. Figure~\ref{line2} shows a slice along a horizontal line  crossing both 1QL and 2QL steps.  
A shift of the $dI/dV$ curves in the direction of the bulk valence band is also present on the steps as well as on a line defect crossed by the scan line.  

On upper terraces the effects observed for $V_0$-  and $V_D$-related quantities are similar.  As our quantitative analysis deals with the shift of the bulk energy structure (see below),  further discussion is given in terms of $V_0$. 

\subsubsection{STS results on a two-dimensional grid}
The second type of  STS measurements performed is scanning tunneling spectroscopy on a two-dimensional grid.  
It provides many more details of the spatial variation of various physical properties in the vicinity of the step.
  
\begin{figure}
\includegraphics[width=0.44\textwidth]{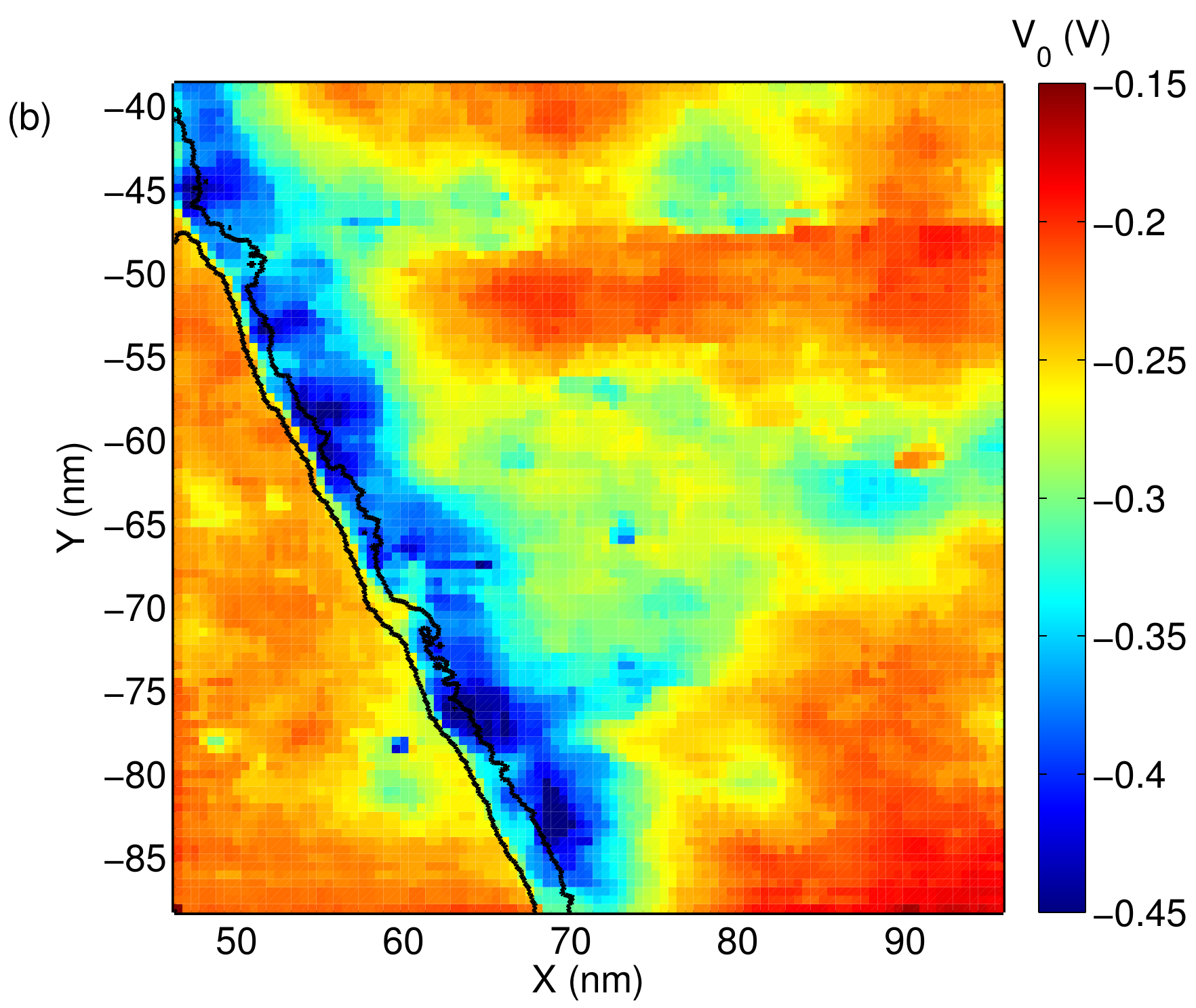}
\includegraphics[width=0.44\textwidth]{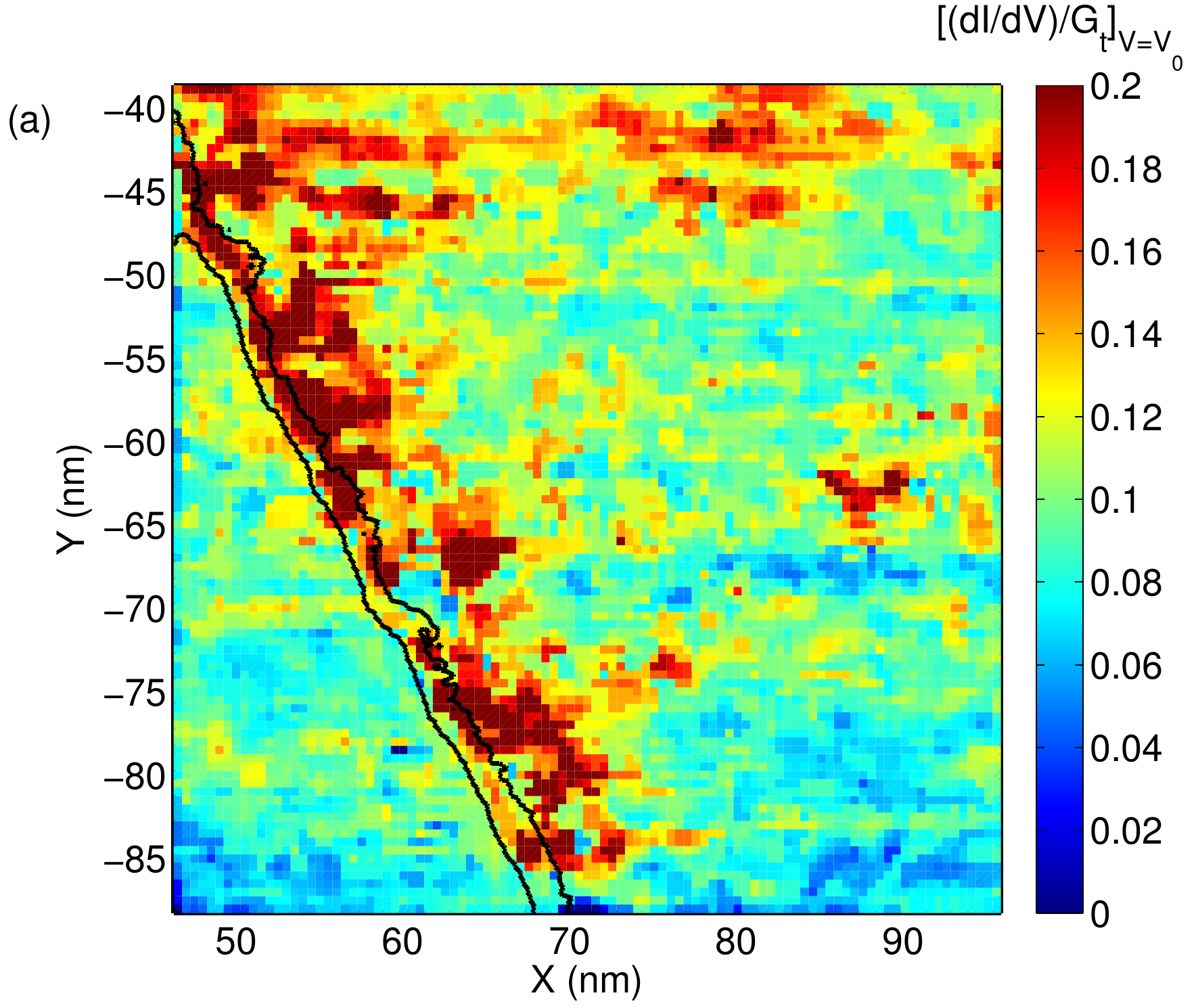}
\caption{Distributions of (a) $V_0$, and (b)
$[(dI/dV)/G_t]_{V=V_0}$  calculated from a $100\times 100$ array of {\em I-V} curves collected on the Bi$_2$Se$_3$  surface shown in Fig.~\protect\ref{surface}(a).  Set point: $V_t = -0.4$~V, $I_t = 100$~pA, $T=5$~K.}
\label{musigma1}
\end{figure}

\begin{figure}
\includegraphics[width=0.44\textwidth]{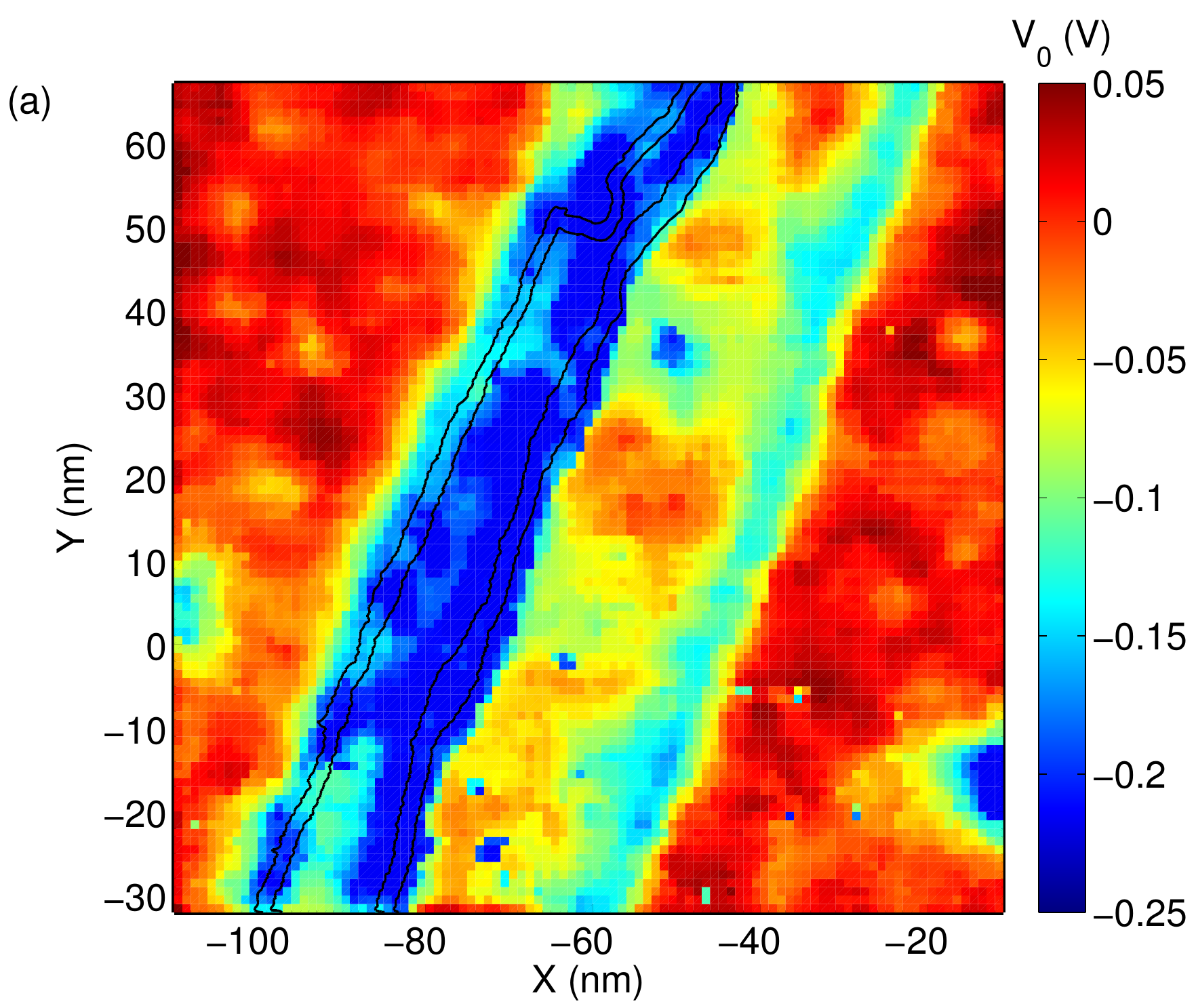}  
\includegraphics[width=0.44\textwidth]{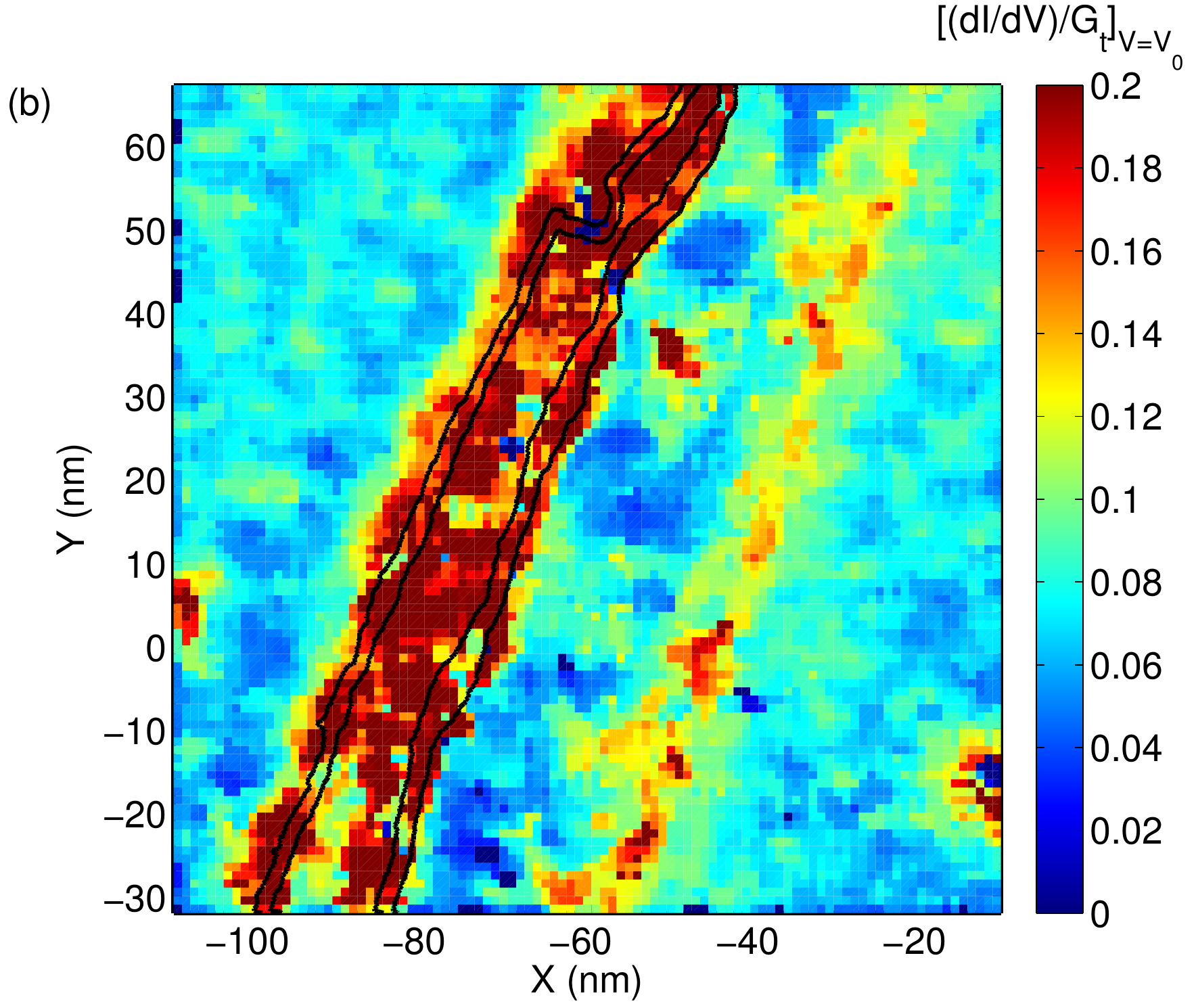}  
\caption{Distributions of (a) $V_0$, and (b)
 $[(dI/dV)/G_t]_{V=V_0}$  on the Bi$_2$Se$_3$  surface shown in Fig.~\protect\ref{surface}(b).   Set point: $V_t = -0.4$~V, $I_t = 100$~pA, $T=5$~K.}
\label{mu2}
\end{figure}

Figures~\ref{musigma1}(a) and \ref{mu2}(a) visualize distributions of $V_0$ over the Bi$_2$Se$_3$ surfaces shown in Fig.~\ref{surface}. $V_0$ exhibits smooth variation within $\pm 0.05$~V over the surface except for the edge step where it decreases by 0.1-0.2~V and reaches its minimum at the step edge. The observed behavior corresponds to that obtained from scans along the lines (Figs.~\ref{line1} and \ref{line2}). In addition, macroscopic defects (point and linear defects) are visualized using this method (Fig.~\ref{mu2}).

Another data set of  interest is the LDOS distribution over the surface. 
The distributions of  $[(dI/dV)/G_t]_{V=V_0}$ for group I and II samples are shown in Figs.~\ref{musigma1}(b) and \ref{mu2}(b) respectively. Clearly, the edge provokes an increase in  $[(dI/dV)/G_t]_{V=V_0}$ within the 10-nm region. 

\section{Discussion}
\begin{figure}
\includegraphics[width=0.4\textwidth]{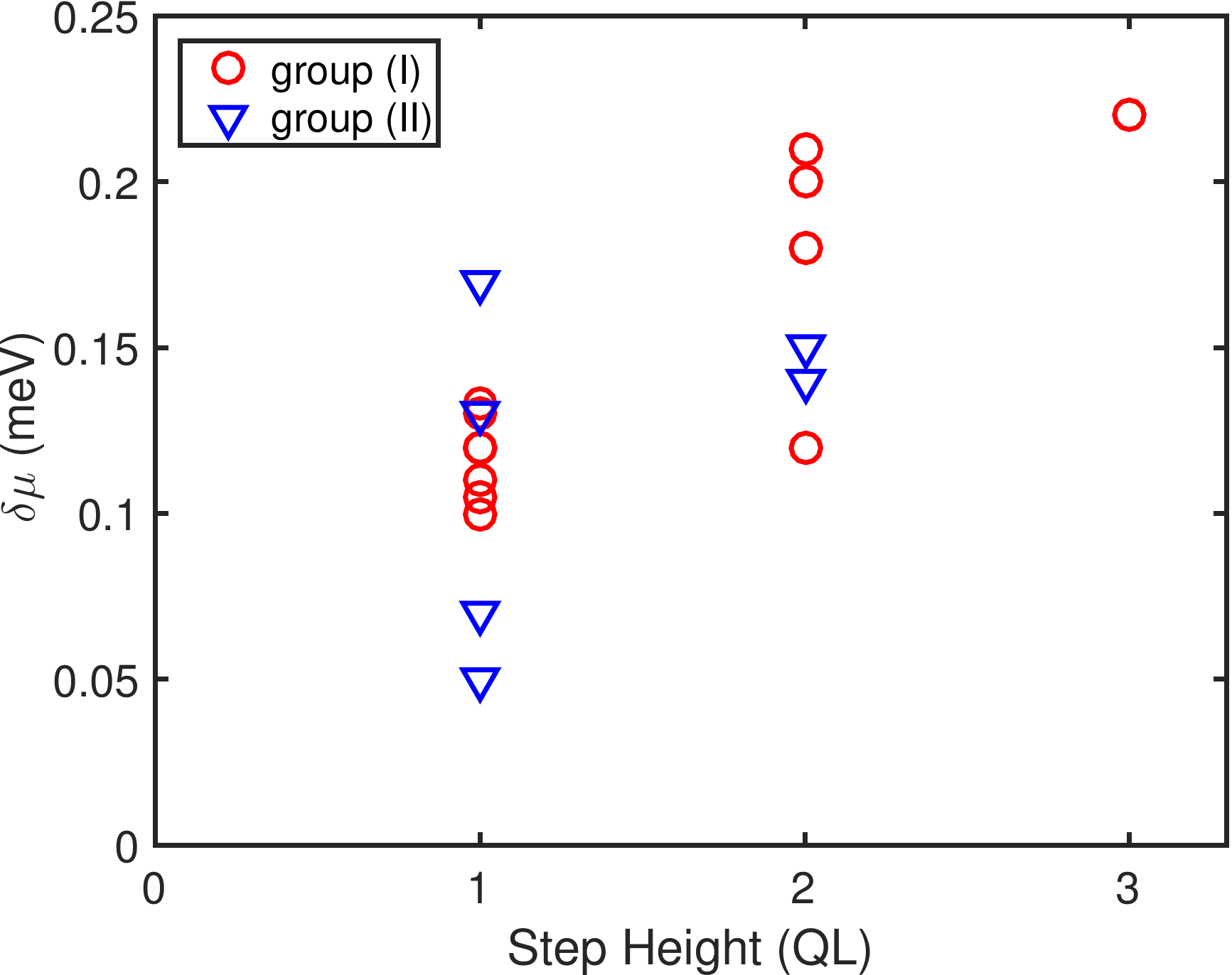}
\caption{Correlation between the step height and the band bending in all studied  samples of groups I (circles) and II (triangles).} \label{qlv}
\end{figure}

The most apparent effect is the band bending in the vicinity of surface steps. The bending always has  the same sign, is directed towards the valence band, varies along a step edge, and depends on the step height. The chemical potential measured from the Dirac point position far from the step edge is  then $\mu=-eV_0$.
The dependence of $\delta \mu$ on  the step height is shown in Fig.~\ref{qlv}, where all the data obtained along individual scan lines are summarized. The correlation between these two quantities is clearly seen: a higher step provokes a larger increase in $\mu$. The typical values are $\delta \mu =0.1\pm 0.05$~eV for 1QL and  $0.15\pm 0.05 $~eV for 2QL steps. The data scattering corresponds to the typical level of $\mu$ fluctuations far from the edge steps and other macroscopic defects. 

The shift of the chemical potential level towards the bulk conduction band corresponds to a reduction of the work function near a step edge. A similar effect is well known for ordinary metals and results from the redistribution of electron wave functions between the outer and inner edges of surface steps  \cite{Smoluchowski}.  
We see two quantitative differences in comparison with ordinary metals such as gold \cite{Jia}:  the value of the effect is smaller (0.1--0.2~eV instead of $0.9\pm 0.3$~eV in gold \cite{Jia}), but the characteristic length of the effect is bigger ($\delta l\sim 10$~nm instead of $0.65\pm 0.01$~nm in gold \cite{Jia}). The smaller value of the shift and bigger length reflect the participation of the surface states with smaller electron density, $n_{3D}$, near the surface ($n_{3D}\approx k_F^2/4\pi\lambda_2 \sim 10^{20}$~cm$^{-3}$ for $1/k_F, \lambda_2\sim 1$~nm) than that in normal metals ($n_ {3D}\sim 10^{22}-10^{23}$~cm$^{-3}$ ). 

\begin{figure}
\includegraphics[width=0.4\textwidth]{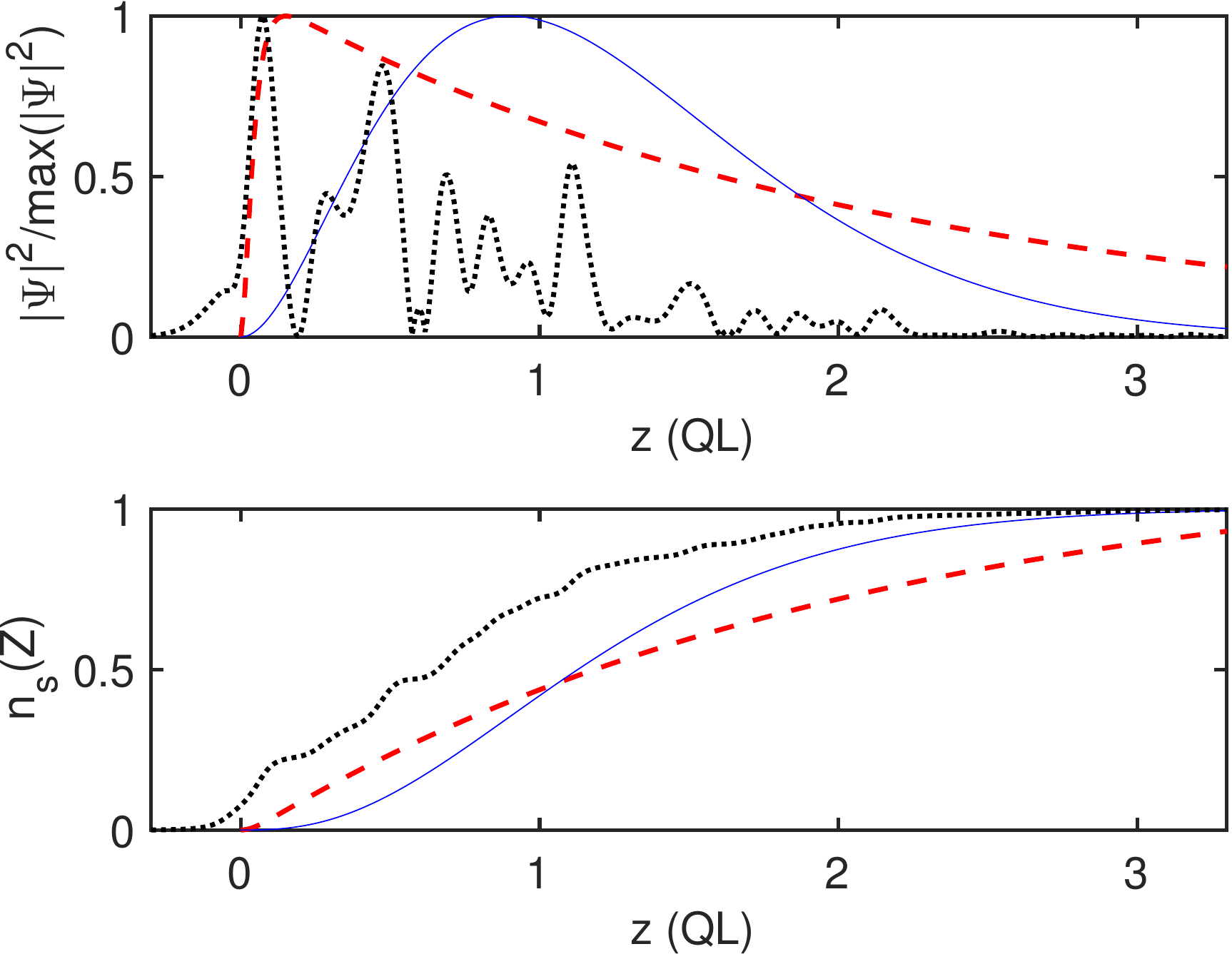}
\caption{(a) Wave-function amplitudes and envelopes normalized by their maxima in various models. Black dotted  line: {\em ab initio} calculations \cite{Eremeev} at $k=0.2$~nm$^{-1}$; red dashed line: Eq.~(\ref{eq}) with the data from Ref.~\cite{Shan} at $k=0.4$~nm$^{-1}$;  blue solid line: Eq.~ (\ref{eq}) with the data from Ref.~\cite{Shan} extracted from {\em ab initio} calculations of Ref.~\cite{BiTeBiSe}  at $k=0.4$~nm$^{-1}$ (a). Respective set of $n_s(Z)\ (b).$} 
\label{wf}
\end{figure}

The value of $\delta\mu$ depends on charges located at the outer and inner edges of the step,  the distance between the charges (i.e. the step height) and the screening length along the step-side surface. Overlapping of the charge localization regions reduces the charge to be screened. The charge localization region is defined by the wave-vector components of filled states.
Analytical expressions for surface state wave function components inside the crystal can be written in the form \cite{Shan}:
\begin{equation}
\Psi=\left[    
\begin{aligned}
  A_1 \\
  A_2 \\
  A_3(k_x,k_y) \\
  A_4(k_x,k_y)
 \end{aligned} 
 \right] (e^{-\lambda_2z}-e^{-\lambda_1z}),
\label{eq}
\end{equation}
where $A_i$ are the wave function components,  $z$ is the distance to the surface (from the bulk side), and $\lambda_{1,2}$ are inverse characteristic distances. 
Figure~\ref{wf}(a) shows the envelopes, $e^{-\lambda_2z}-e^{-\lambda_1z}$, and $|\Psi|^2$ obtained under various approximations for Bi$_2$Se$_3$ \cite{Eremeev,Shan}. 
The nominal fraction of helical electronlike states affected by a step of height $Z$ can be estimated from 
\begin{equation}
n_s(Z)\propto \int_{0}^{Z} \int_{0} ^{k_y(\mu)}\int_{0}^{k_x(\mu)} | \Psi(k_x, k_y, z)|^2 dk_x dk_y dz.
\label{Qz}
\end{equation} 
Figure~\ref{wf}(b) shows the $n_s(Z)$ dependence obtained for different models. We see that in all cases the surface states are not limited by the first QL: a noticeable fraction extends into the second QL. We expect therefore that the charge  located at the 1-QL step edge is smaller than that for the 2-QL one.  

Other mechanisms may also contribute to this shift. 
A contribution may come from the dependence of the Dirac point energy position on the facet orientation.  In particular, the position of the Dirac point on  the (111) surface is $\sim 0.15$~eV lower than  its position on  side surfaces \cite{Lee,Moon,Virk}, so band bending is expected.  A trace of the respective change of the Dirac point position on a side surface is seen in Figs.~\ref{line1} and \ref{line2} as a deflection of  $V_D$ from $V_0$ in the transition region between two terraces. A contribution to  the chemical potential shift may also come from a difference in work functions $W$ of different surfaces of Bi$_2$Se$_3$. The first-principle calculation \cite{Lee} gives values of the  work function $W_{(111)} = 5.84$~eV and $W_{(\overline 110)} = 5.04$~eV for the unrelaxed $(111)$  and $(\overline 110)$ surfaces respectively, and 5.8 and~4.97 eV for relaxed ones. Only a small fraction $\sim \Delta Z/\delta l\approx 0.1$ of the work function difference works in this case because of small step height $\Delta Z$. The sign of the work function change corresponds to our observations.  In addition, a charge accumulated by dangling bonds may also contribute to this potential difference.

Screening in topological insulator surface states was considered in Ref.~\cite{Adam}. The Tomas-Fermi analysis gives the dielectric function $\epsilon(q)=1+q_{TF}/q$, where $q_{TF}=\eta r_s k_F$, $\eta\approx 1$ in our case, $r_s=e^2/\kappa \hbar v_F$ is the interaction parameter, and  $\kappa$ is the dielectric constant. Taking $r_s=0.1$ ~\cite{Adam} and $k_F=1$~nm$^{-1}$ one gets the screening length $\sim 1/q_{TF}\approx 10$~nm, in agreement with our data.

 The most essential  question is whether the observed increase in  $[(dI/dV)/G_t]_{V=V_0}$ really corresponds to the increase in  LDOS. It will be argued below that a major part (if not all) of the observed  $[(dI/dV)/G_t]_{V=V_0}$  increase is actually caused by a modification of the transmittance of the vacuum tunneling barrier due to a shift of the chemical potential level. 

Let us consider the effect of the chemical potential shift  $\delta \mu$ on the tunneling spectra $dI/dV$ following Ref.~\cite{Fedotov}.  For our purposes it is enough to analyze the simplest model 
in the zero-temperature limit.
The tunneling current can be written as 
\begin{equation}
I(V)=A\int\limits_{0}^{V}\rho_s(E)T(E,V)\rho_t(E-eV)dE,
\label{IV}
\end{equation}
where $\rho_s$ and $\rho_t$ are, respectively, surface and tunneling tip densities of states, and $T(E,V)$ is the transmittance of the tunneling gap.
\begin{figure}
\includegraphics[width=0.23\textwidth]{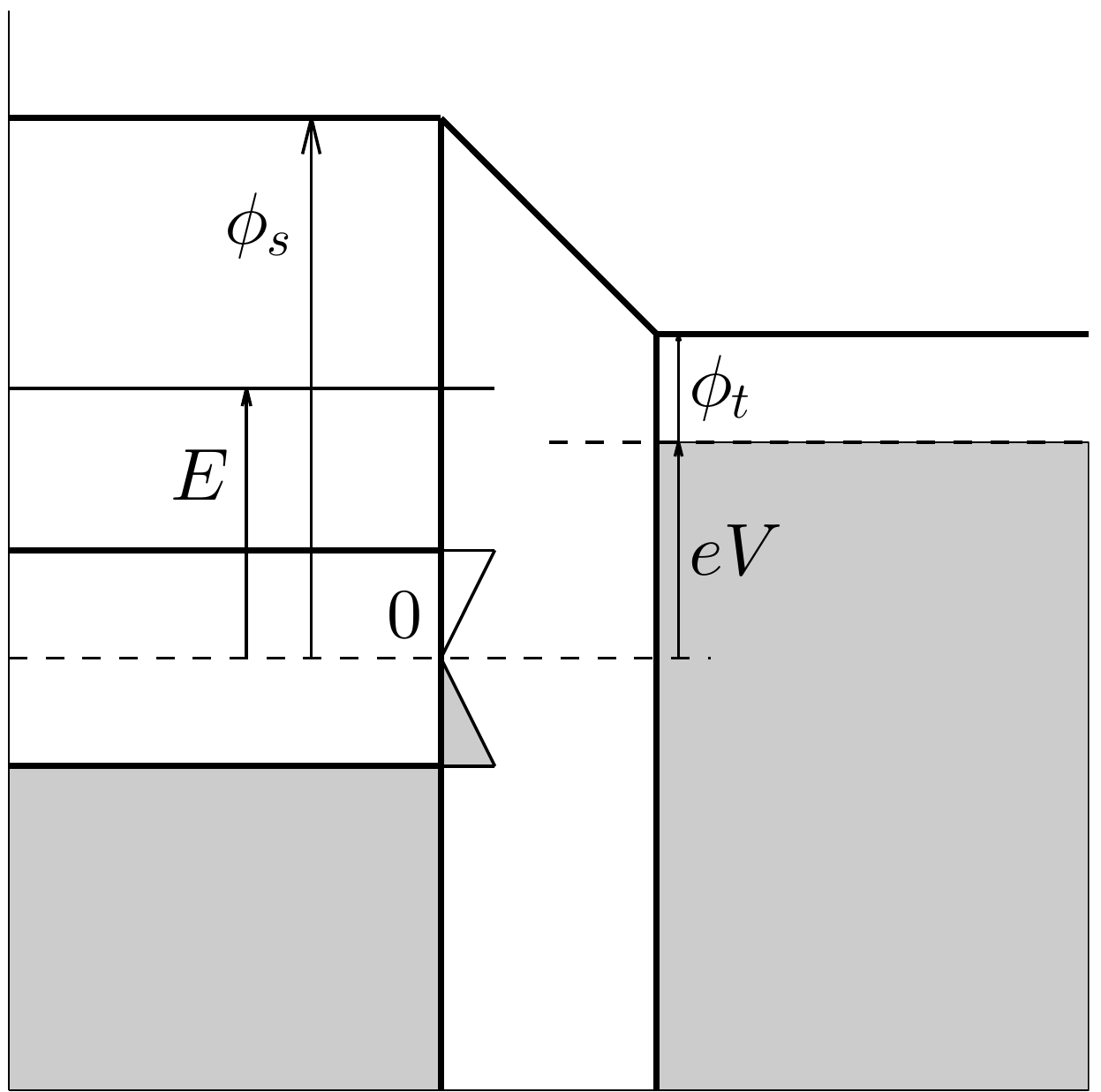}
\includegraphics[width=0.23\textwidth]{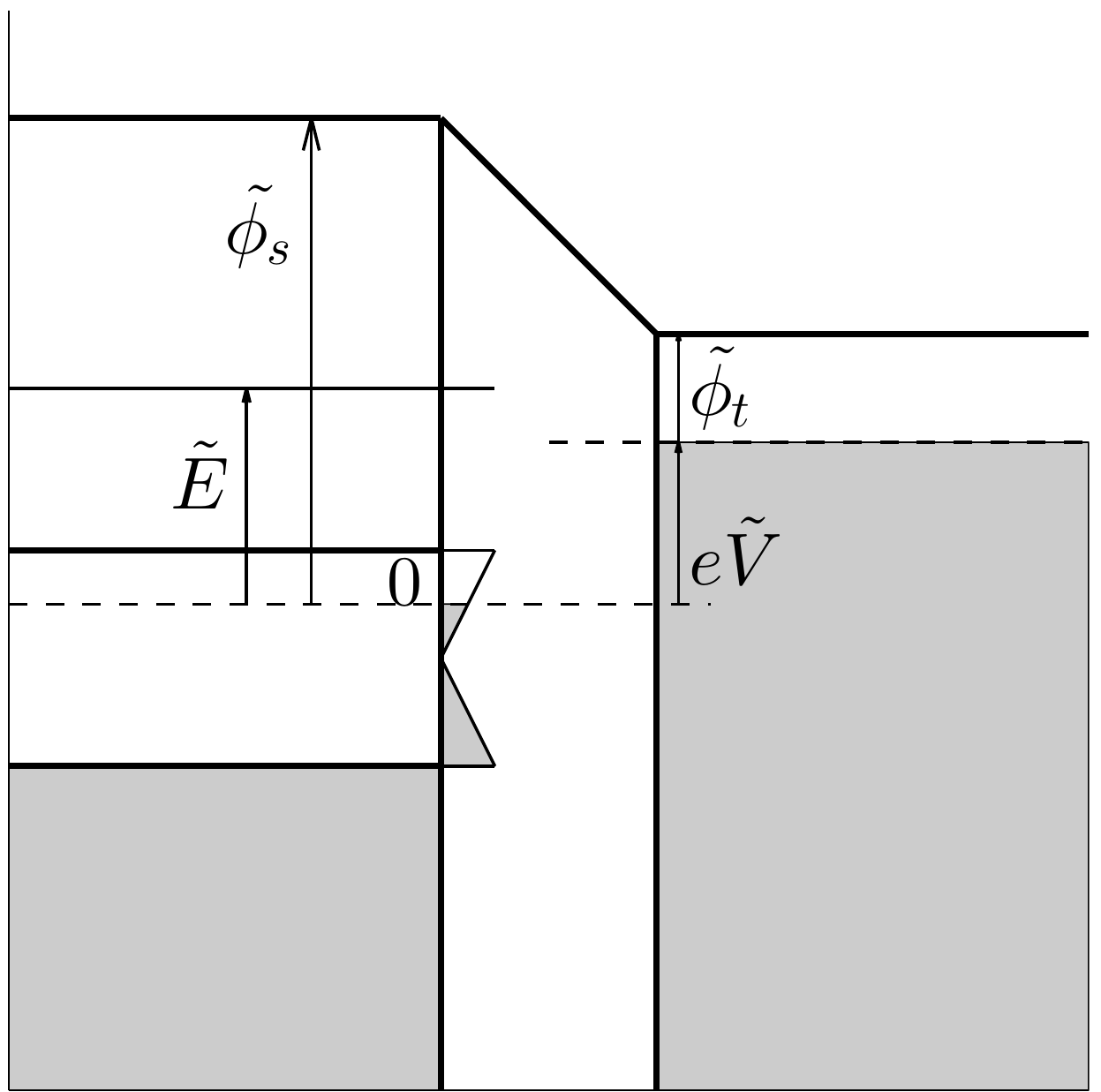}
\caption{Energy diagrams for different chemical potential positions. Bulk DOS is omitted for simplicity.}
\label{model}
\end{figure}
The shift of the chemical potential modifies both $\rho_s$ and $T$ 
(see Fig.~\ref{model}), the differential tunneling conductance at the Dirac point, $G=dI/dV|_{V=V_D}$, is then \cite{Fedotov}
\begin{equation}
\begin{aligned}
G_D(\delta \mu)=G_{D}(0)+\\
+eA\int_{0}^{\delta\mu}\rho_s(E)\frac{d\rho_t(E-eV)}{dE}T(E,V,z)dE-\\
-A\int_{0}^{\delta\mu}\rho_s(E)\rho_t(E-eV)\frac{\partial T}{\partial V}(E,V,z)dE.
\label{eq:deltaG}
\end{aligned}
\end{equation}
We see that there is a correction to $G_D(0)$. This correction is most noticeable if $G_D$ is small initially, i.e. when $\mu=eV_D$ or very close to this value, as it takes place in the group II samples.
For numerical simulations we use the transmittance in the Wentzel-Kramers-Brillouin approximation,
\begin{equation}
T(E,V)\approx\exp\left( -z\frac{2\sqrt{2m}}{\hbar}\sqrt{\phi+\frac{eV}{2}  -E }    \right),
\label{T}
\end{equation}
where $z$ is the tip-sample distance and $\phi=(\phi_s+\phi_t)/2$ is the mean work function of the sample surface and the  tip. 

\begin{figure}
\includegraphics[width=0.4\textwidth]{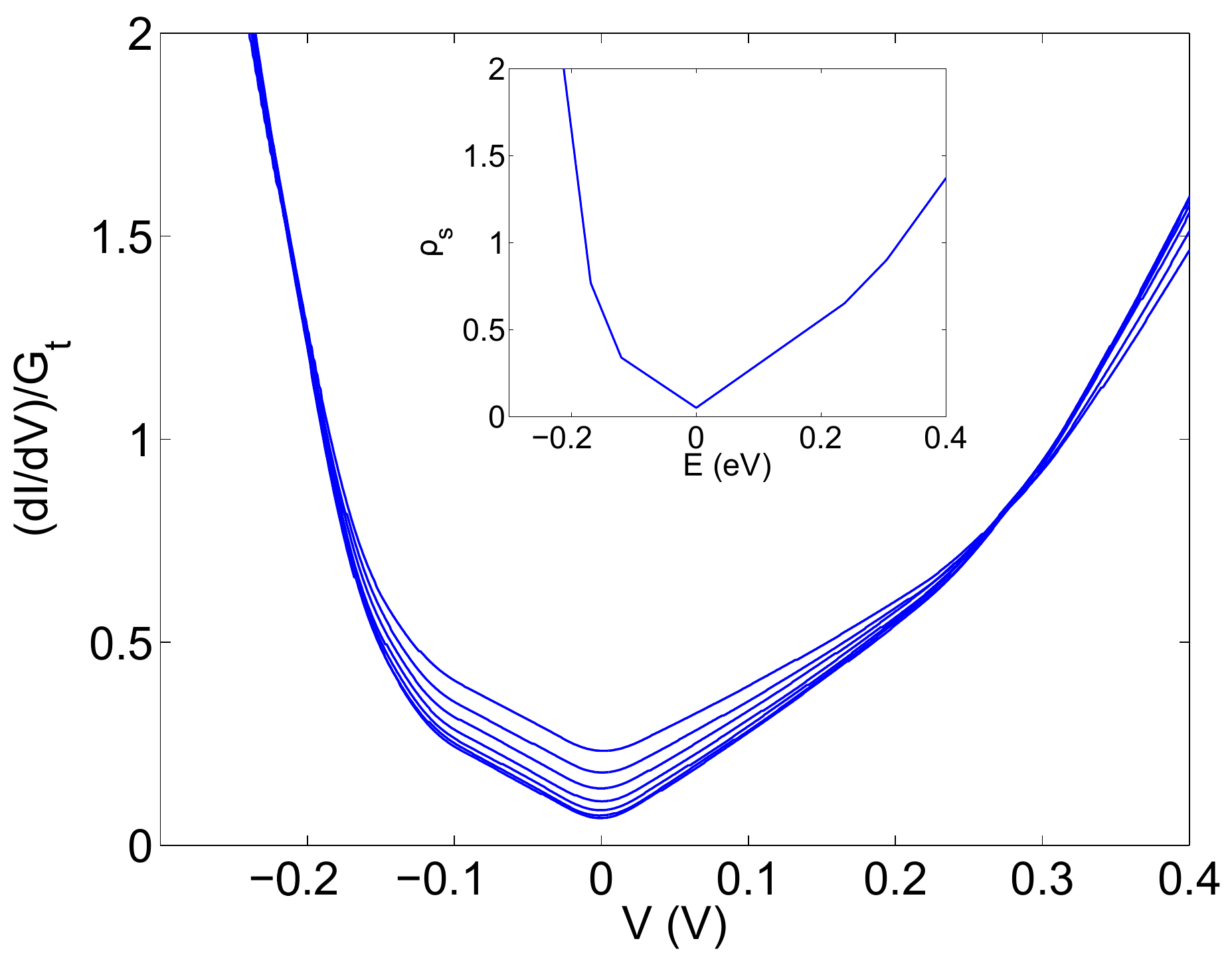}
\caption{A set of $dI/dV$ curves obtained in accordance with Eqs.~(\ref{IV}) and (\ref{T})  for the same energy spectra $\rho_s$ with different positions of the chemical potential level measured from the Dirac point $\mu = 0, 0.05,  0,1, 0.15, 0,2, 0.25, 0.3 $~eV, $\phi_s=5.5$~eV, $\phi_t=5.0$~eV, $z=1$~nm.}
\label{didv}
\end{figure}
\begin{figure}
\includegraphics[width=0.45\textwidth]{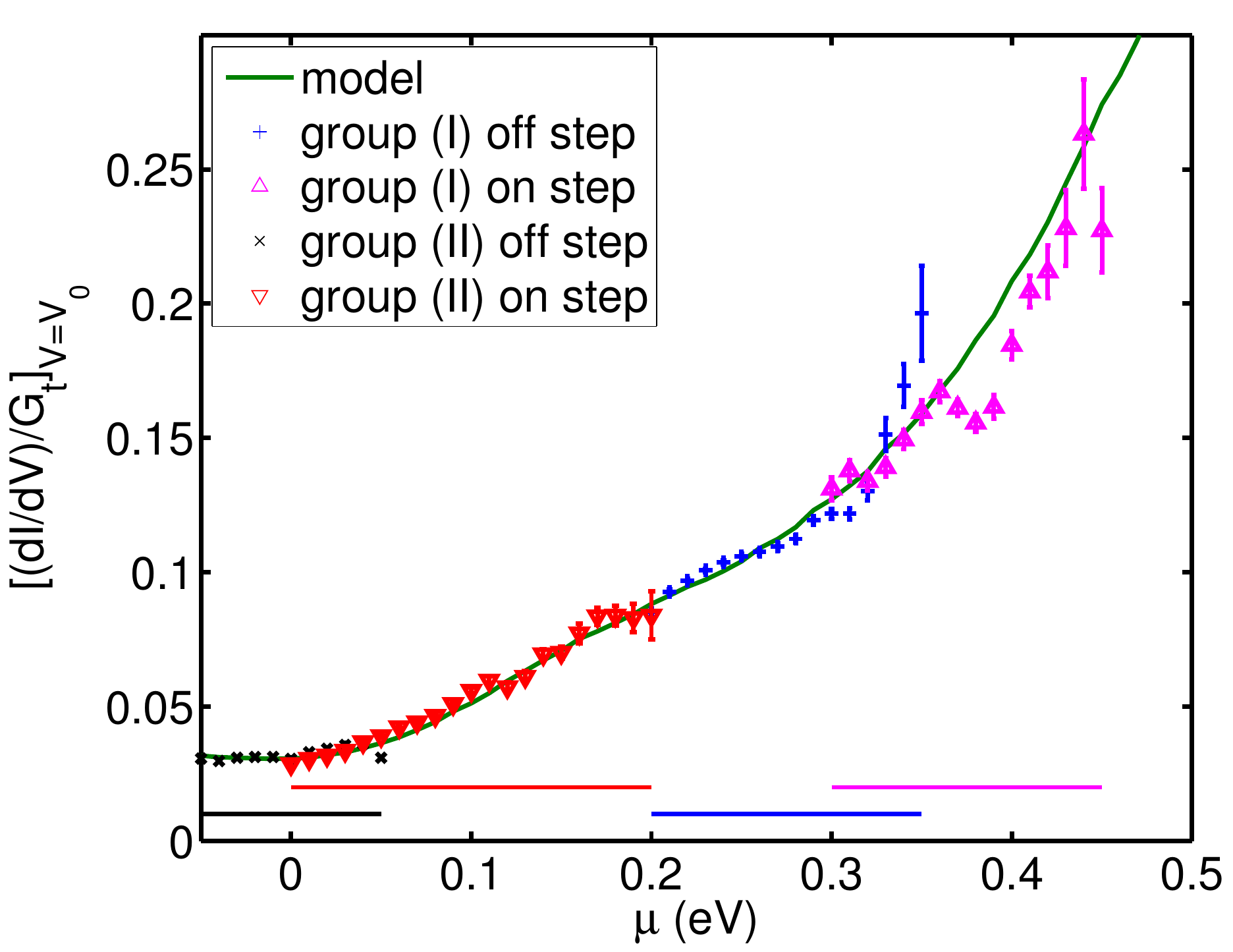}
\caption{Effect of the chemical potential shift on $[(dI/dV)/G_t]_{V=V_0}$. The solid line corresponds to the model density of states shown in the inset in  Fig.~\ref{didv}. Experimental points correspond to the data in Figs.~\ref{musigma1} and \ref{mu2}. See text for details. Horizontal lines show the chemical potential variation range of the respective dataset.}
\label{LDOS}
\end{figure}

Figure~\ref{didv} shows a set of simulated normalized $dI/dV$ curves obtained for the same $\rho_s(E)$ (see inset) at different values of $\delta\mu$ in accordance with Eqs.~(\ref{IV}) and (\ref{T}). The model density of states, $\rho_s$, is chosen to produce $dI/dV$ curves similar to the measured ones [Figs.~\ref{vah}(a),(b)].
We see that the typical shift of the chemical potential by 0.15~eV results in approximately the same increase in $[(dI/dV)/G_t]_{V=V_0}$ by a factor of 2-3 as observed experimentally [Figs.~\ref{line1}(d), and \ref{line2}(d)]. Note that a similar effect is also present in STS data collected far from defects: $[(dI/dV)/G_t]_{V=V_0}$ in group II samples is apparently deeper than that in group I samples (Fig.~\ref{vah}, and Fig.~\ref{musigma1}(b), \ref{mu2}(b) \footnote{Bear in mind the difference in normalization of $dI/dV$ curves of groups I and II samples.}). Similar behavior can also be  found in Ref.~\cite{Dai}. 

To test the presence of edge states we compare the effects of the chemical potential shift near and far from the steps. For this purpose we select 5~nm wide stripes near the steps and  regions 20~nm away from the edges on the upper terraces for samples shown in Fig.~\ref{surface}. As $V_0$ is locked to the bulk energy structure, $[(dI/dV)/G_t]_{V=V_0}$ will be used  for comparison with the model described above. We plot $[(dI/dV)/G_t]_{V=V_0}$   {\em vs.} $\mu=-eV_0$ by using the data shown in Figs.~\ref{musigma1} and \ref{mu2}. As the number of data points is very large, they are averages over 20 meV intervals, and margins of error are calculated accordingly. The results are shown in Fig.~\ref{LDOS} \cite{comment1}.  The data points form four overlapping regions. The results form a unique curve without any noticeable discontinuity between the different regions. Moreover, this curve can be almost perfectly fitted by Eqs. (\ref{IV}) and (\ref{T})  with the model density shown in the inset in Fig.~\ref{didv} if 20\%  variation of $\rho_t(E)$ is taken into account \cite{comment}.  As $[(dI/dV)/G_t]_{V=V_D}\leq [(dI/dV)/G_t]_{V=V_0}$, so no extra contribution from edge states is  present in either $V_0$ or i$V_D$.

Finally, the observed increase in normalized $dI/dV$ near the step edge on the surface of the topological insulator Bi$_2$Se$_3$ is practically totally accounted for by the effect of the  shift of the chemical potential level.

\section{Conclusion}
We demonstrated  that the energy structure of the surface states of the topological insulator  Bi$_2$Se$_3$  revealed by STS exhibits dramatic changes near the step edge: there is a shift in the chemical potential level which is accompanied by an apparent increase in the normalized $dI/dV$ value at the Dirac point. Various contributions to the chemical potential shift were analyzed. The most probable one corresponds to a reduction in the work function on stepped surfaces of topologically trivial  metals. Quantitative differences (a smaller value of the chemical potential shift and bigger spatial scale) reflect features of topological insulators: smaller surface current carrier concentration and smaller wave vectors. We also demonstrated  that the apparent increase in  normalized $dI/dV$ near the step edge is actually an artifact of the STS method. This increase is practically entirely accounted for by the voltage-dependent transparency of the tunneling barrier and therefore cannot be considered an indication of the increase in the LDOS near the step edge. 

\begin{acknowledgements}
We are grateful to V.A. Sablikov and V.V. Pavlovskii for useful discussions, V.F. Nasretdinova for help in crystal growth, and S.V. Eremeev for useful discussion and providing the data fromf Ref.~\cite{Eremeev}.  Financial support from RSF (Project \# 16-12-10335, experimental part) and RFBR (Hroject \#16-02-00677, effect of the chemical potential) is acknowledged.
\end{acknowledgements}


\begin{thebibliography}{99}
\bibitem{comment1}As the data shown in Figs.~\ref{musigma1} and \ref{mu2} were obtained with different $\Delta V$, we used the empirical factor $\beta=G_t|_{\Delta V={\rm ~0.5 eV}}/G_t|_{\Delta V=0.7{\rm  ~eV}}$ for rescaling $[(dI/dV)/G_t]_{V=V_0}$ data of group II samples.  The value $\beta=0.4251$ is derived from the I-V curve shown in Fig.~\ref{vah}(a).
\end{thebibliography}
\end{document}